\documentclass[doublecol]{epl2}

\usepackage{upgreek}
\usepackage{amsmath}
\usepackage{txfonts}
\usepackage{booktabs}
\usepackage{color}
\usepackage{array}
\usepackage{graphicx}
\usepackage{epstopdf}
\usepackage{bm}
\usepackage{cmap}
\usepackage{tabularx}
\usepackage{multirow}
\usepackage{ulem}

\def\lesssim{\;\raise0.3ex\hbox{$<$\kern-0.75em\raise-1.1ex\hbox{$\sim$}}\;}
\def\gtrsim{\;\raise0.3ex\hbox{$>$\kern-0.75em\raise-1.1ex\hbox{$\sim$}}\;}

\newcommand{\partDer}[2]{\frac{\partial #1}{\partial #2}}

\newcommand{\diverg}{\mathrm{div}}
\newcommand{\pF}{p_{\mathrm{F}}}

\newcommand{\gcc}{\mathrm{g~cm}^{-3}}

\newcommand{\xHall}{x_{\mathrm{H}}}

\revision

\title{Shear viscosity in magnetized neutron star crust}
\shorttitle{Shear viscosity in neutron star crust}

\author{D. D. Ofengeim\inst{1,2} \and D. G. Yakovlev\inst{1}}
\shortauthor{D. Ofengeim \and D. Yakovlev}

\institute{
  \inst{1} Ioffe Institute,
Politechnicheskaya 26, 194021, St.\ Petersburg, Russia\\
  \inst{2} St.~Petersburg Academic University,  Khlopina Street 8/3,
St.~Petersburg 194021, Russia }


\abstract{The electron shear viscosity due to Coulomb scattering of
degenerate electrons by atomic nuclei throughout a magnetized
neutron star crust is calculated. The theory is based on the shear
viscosity coefficient calculated neglecting magnetic fields but
taking into account gaseous, liquid and solid states of atomic
nuclei, multiphonon scattering processes, and finite sizes of the
nuclei albeit neglecting the effects of electron band
structure. The effects of strong magnetic fields are included in
the relaxation time approximation with the effective electron
relaxation time taken from the field-free theory. The viscosity in a
magnetized matter is described by five shear viscosity coefficients.
They are calculated and their dependence on the magnetic field and
other parameters of dense matter is analyzed. Possible applications
and open problems are outlined.}

\begin{document}
\maketitle

\section{Introduction}

Shear viscosity is important in
neutron stars. It regulates dissipation of hydrodynamical motions
in these stars, for instance,  relaxation of differential rotation
to a rigid-body rotation,
damping of various waves and oscillations. In particular,
it can damp instabilities associated with the emission of
gravitational waves (e.g., r-mode instability
\cite{RMODES01,HASKELL15}) which is important
for planning advanced gravitational wave experiments.

The shear viscosity has been studied in
different layers of neutron stars, in the crust and the core, but
neglecting the effects of the magnetic fields. However,
neutron stars can possess strong magnetic fields \cite{MAGNETARS}.
Typical surface fields $B$ of ordinary neutron
stars (e.g., radio pulsars) can be as high as $10^{12}-10^{13}$ G
\cite{MANCHESTER2005},
whereas the surface fields of highly magnetized
neutron stars (magnetars) 
can be more than
one order of magnitude larger \cite{OlKas2014}. The internal
fields can be even stronger.
To the best of our
knowledge, the only calculation of the shear viscosity in a magnetic
field related to compact stars was done by Haensel and
Jerzak \cite{HaensJer1989} for
strange quark stars.

We consider the electron shear viscosity in a magnetized neutron
star crust. The electrons are important momentum carriers
there\cite{FI1976}; the basic electron scattering mechanism is the
Coulomb scattering off atomic nuclei. The electrons mainly
constitute a strongly degenerate relativistic and weakly interacting
gas. The viscosity in the field-free case was analyzed by Flowers
and Itoh \cite{FI1976} and detailed in
\cite{Pethick,ChugYak2005,SHTERN2008}

Our analysis is based on the calculations \cite{ChugYak2005}
of the electron shear
viscosity in the neutron star crust neglecting
the magnetic field. The field makes the electron transport
anisotropic which is well studied for the cases of electron
electric and thermal conductivities \cite{PPP2015}. Here we
investigate the effect of magnetic fields on the shear viscosity.

\section{Formalism}
\label{s:Formalism}

The electron distribution function is taken in the form
\begin{equation}
\label{eq:f=f0+df}
f(\bm{p}) = f_0(\epsilon) + \delta f(\bm{p}),
\end{equation}
where $\epsilon$ and $\bm{p}$ are, respectively, the electron
energy (with the rest-mass contribution) and momentum; $f_0(\epsilon)$
is the equilibrium Fermi-Dirac distribution, and
$\delta f(\bm{p})$ is a nonequilibrium correction.

The calculation of $\delta f(\bm{p})$ is based on the linearized
Boltzmann equation within the relaxation time approximation of the
collision integral. This equation can be written as
\begin{equation}
\label{eq:boltzman}
\left( \partDer{f_0}{\mu} \right)\left( v_{\alpha} p_{\beta}
\partDer{V_{\alpha}}{x_{\beta}} - \frac{1}{3}v_{\alpha}p_{\alpha}
\diverg \bm{V} \right) = -\frac{\delta f}{\tau} + \frac{e}{c}\, (
\bm{v} \times \bm{B})
\partDer{\delta f}{\bm{p}},
\end{equation}
where $\bm{B}$ is the magnetic field, $\mu$ is the electron chemical
potential, $\bm{v}$ the electron velocity, and
\begin{equation}
\label{eq:Vdef}
V_{\alpha\beta} = \frac{1}{2} \left(
\partDer{V_{\alpha}}{x_{\beta}} + \partDer{V_{\beta}}{x_{\alpha}}
\right),
\end{equation}
$\bm{V}$ being the velocity of matter elements (assumed to be small,
particularly, non-relativistic). Furthermore, $e=|e|$ is the
elementary charge, $c$ is light speed, and $\tau=\tau(\epsilon)$ is
an effective electron relaxation time. Equation (\ref{eq:boltzman})
is similar to that for a non-relativistic non-degenerate plasma
\cite{LL_X} where one should replace $f_0/(k_{\rm B}T) \to
\partial f_0 / \partial \mu$ ($k_{\rm B}$ being the Boltzmann
constant). Since we study strongly degenerate electrons, it is
sufficient to set $\epsilon = \mu$ in all functions of $\epsilon$
which vary weakly within the thermal width of the Fermi level,
$|\epsilon - \mu| \lesssim k_{\rm B}T$. In addition, one can set
$\partial f_0 / \partial \mu \to \delta(\mu-\epsilon)$. Similar
equation was written and solved for degenerate ultra-relativistic
quark plasma \cite{HaensJer1989}. A simple analysis show that it is
valid for degenerate electrons of any degree of relativity. It is
straightforward to use  \cite{HaensJer1989} and obtain the results
required for our case.

A viscous stress tensor in the restframe of
the matter is
\begin{equation}
\label{eq:sigmaDef}
\sigma_{\alpha\beta} = -
\frac{2}{(2\pi \hbar)^3}
\int {\rm d} \bm{p}\, v_{\alpha} p_{\beta}\, \delta f(\bm{p}).
\end{equation}

Using the results of \cite{LL_X} one obtains $\sigma_{\alpha\beta} =
\sum_{i=0}^4 \eta_i g_{i\,\alpha\beta}$, where
$\eta_0,\ldots,\eta_4$ are five coefficients of shear viscosity in a
magnetic field, and
\begin{eqnarray}
g_{0\,\alpha\beta} & = & \left( 3b_{\alpha}b_{\beta} -
\delta_{\alpha\beta} \right)\left(
V_{\gamma\delta}b_{\gamma}b_{\delta} - \frac{1}{3}\,\diverg\bm{V}
\right),
\nonumber \\
g_{1\,\alpha\beta} & = & 2V_{\alpha\beta} - \delta_{\alpha\beta}
\,\diverg\bm{V} - 2V_{\alpha\gamma} b_{\beta}b_{\gamma} -
2b_{\alpha}
V_{\beta\gamma}b_\gamma +\nonumber\\
&   & + \delta_{\alpha\beta}V_{\gamma\delta}b_{\gamma}b_{\delta} +
b_{\alpha}b_{\beta}\,\diverg\bm{V}
 + b_{\alpha}b_{\beta}V_{\gamma\delta}b_{\gamma}b_{\delta},
\nonumber \\
g_{2\,\alpha\beta} & = & 2\left( V_{\alpha\gamma}b_{\beta}b_{\gamma}
+ b_{\alpha}V_{\beta\gamma}b_{\gamma} -
2b_{\alpha}b_{\beta}V_{\gamma\delta}b_{\gamma}b_{\delta} \right),
\nonumber \\
g_{3\,\alpha\beta} & = & - V_{\alpha\gamma}b_{\beta\gamma} -
b_{\alpha\gamma}V_{\beta\gamma} +
b_{\alpha\gamma}b_{\beta}V_{\gamma\delta}b_{\delta}
 +  b_{\alpha}b_{\beta\gamma}V_{\gamma\delta}b_{\delta},
\nonumber \\
g_{4\,\alpha\beta} & = & - 2\left(
b_{\alpha\gamma}b_{\beta}V_{\gamma\delta}b_{\delta} +
b_{\alpha}b_{\beta\gamma}V_{\gamma\delta}b_{\delta} \right).
\label{eq:sigmaFin}
\end{eqnarray}
In this case $\bm{b}$ is a unit vector along $\bm{B}$, $\delta_{\alpha \beta}$ is
Kronecker's delta,
$b_{\alpha\beta}=\varepsilon_{\alpha\beta\gamma}b_\gamma$, and
$\varepsilon_{\alpha\beta\gamma}$ is the Levi-Civita tensor.

Making use of the results of \cite{HaensJer1989} for degenerate
electrons of any degree of relativity we have
\begin{eqnarray}
\eta_0 = \frac{ c^2 \pF^5\tau}{15\pi^2\hbar^3 \mu}, \quad \eta_1 =
\eta_0 \, \frac{1}{1+4\xHall^2}, \quad \eta_2 = \eta_0\,
\frac{1}{1+\xHall^2},
\nonumber \\
\eta_3 = \eta_0\, \frac{2\xHall}{1+4\xHall^2}, \qquad \eta_4 =
\eta_0 \, \frac{\xHall}{1+\xHall^2}. \label{eq:eta01234}
\end{eqnarray}
Here $\pF$ is the electron Fermi momentum, $\xHall = \omega \tau$ is
a dimensionless Hall parameter and $\omega=eBc/\epsilon=eBc/\mu$ is
the electron gyrofrequency. In the ultrarelativistic limit ($\mu
\approx \pF c$) these equations coincide with those derived in
\cite{HaensJer1989} (although $\eta_2$ in \cite{HaensJer1989}
contains a typo, an extra factor 4 in the denominator).

Since $\sigma_{\alpha\alpha} = 0$, all the five coefficients are,
indeed, shear viscosities. The behavior of $\sigma_{\alpha \beta}$
under the interchange of indices $\alpha$ and $\beta$ and
tranformation $\bm{B} \to -\bm{B}$ is in line with the Onsager
relations \cite{LL_X}.

\begin{figure}
\includegraphics[width=\columnwidth]{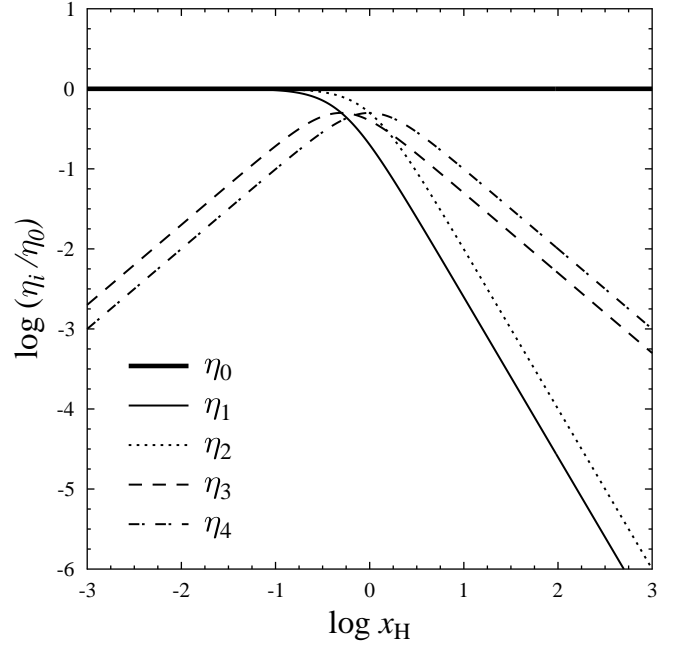}
\caption{\label{fig:lgx-lgeta} Electron shear viscosity coefficients
$\eta_i$ ($i=0,\dots,4$) in units of the field-free viscosity
$\eta_0$ versus the Hall parameter $\xHall=\omega \tau$.}
\end{figure}

The viscous (collisional) energy dissipation rate
[erg~cm$^{-3}$s$^{-1}$] in a shear flow is
given by $T\dot{s}_{\rm coll}=\sigma_{\alpha \beta }V_{\alpha \beta}$,
where $s$ is a specific entropy. In our case
\begin{eqnarray}
   T\dot{s}_{\rm coll}&=&  3 \eta_0\,
    \left( V_{\alpha\beta}b_\alpha b_\beta
    -{ 1 \over 3}\,\diverg\bm{V} \right)^2
\nonumber \\
   &+& \eta_1 \, \left[ 2 V_{\alpha\beta}   V_{\alpha\beta}-
    (\diverg\bm{V})^2 -4 V_{\alpha\beta}V_{\alpha\gamma}
    b_{\beta}b_{\gamma} \right.
\nonumber \\
  &+& \left. 2V_{\alpha\beta}b_{\alpha}b_{\beta}\,\diverg\bm{V}
        +(V_{\alpha\beta}b_{\alpha}b_{\beta})^2 \right]
\nonumber \\
   &+& 4\eta_2 \left[ V_{\alpha\beta}V_{\alpha\gamma}
    b_{\beta}b_{\gamma}
        - (V_{\alpha\beta}b_{\alpha}b_{\beta})^2    \right].
\label{e:heat}
\end{eqnarray}

The nature of
the viscosities $\eta_0 - \eta_4$ is related to the orthogonality of
the tensors $g_0,\ldots,g_4$ to the `magnetic' tensor $b_{\alpha}b_{\beta}$. One has $g_{i\,\alpha\beta} b_{\beta} =
b_{\alpha}g_{i\,\alpha\beta} = 0$ for $i$=1 and 3, and
$g_{i\,\alpha\beta} b_{\alpha}b_{\beta} = 0$ for $i$ =2 and 4, but
\begin{equation}
g_{0\,\alpha\beta}b_{\alpha}b_{\beta} =
2V_{\alpha\beta}b_{\alpha}b_{\beta} -
\frac{2}{3}\,\diverg\bm{V} \ne 0.
\end{equation}
Accordingly, $g_0$ is `parallel' to $b_{\alpha}b_{\beta}$,
$g_2$ and $g_4$ are `transverse' to $b_{\alpha}b_{\beta}$,
whereas $g_1$ and $g_3$ are `fully transverse' to
$b_{\alpha}b_{\beta}$. Since $g_3$ and $g_4$  include $b_{\alpha\beta}$,
they are pseudotensors and
describe Hall-like momentum transfer.

Therefore, in analogy with the well-known classification
of electric and thermal conductivities in a magnetized plasma,
we can call $\eta_0$ the {\it longitudinal viscosity}
(with respect to $\bm{B}$),
$\eta_1$ the {\it fully transverse viscosity},
$\eta_2$ the {\it ordinary transverse viscosity},
$\eta_3$ the  {\it fully Hall viscosity}
and $\eta_4$ the {\it ordinary Hall viscosity}.

One has $\eta_1(\xHall)=\eta_2(2 \xHall)$ and
$\eta_3(\xHall)=\eta_4(2 \xHall)$. These relations become
clear after writing the system of equations for $\eta_0,\ldots,\eta_4$
in a local reference frame with the $z$ axis along $\bm{B}$.
Such a system splits into three subsystems. The first subsystem
is for $\eta_0$ which appears to be determined by $V_{zz}$. The
`driving force' $V_{zz}$ is directed along $\bm{B}$ and does not interfere
with the Lorentz force acting on electrons.
Accordingly, $\eta_0$ is formally
independent of $B$. The second subsystem is for $\eta_2$ and
$\eta_4$ with the `driving force' created by $V_{xz}$ and
$V_{yz}$ (partly along $\bm{B}$ but partly across $\bm{B}$, interferes
with ordinary electron gyrorotation with frequency $\omega$).
This subsystem gives us the ordinary transverse and Hall
viscosities. Finally, the third subsystem is for
$\eta_1$ and $\eta_3$ with the `driving force' created by $V_{xx}$,
$V_{xy}$ and $V_{yy}$. Such a force is fully transverse to $\bm{B}$ and
interferes with the second harmonic of gyrorotation
(with frequency $2 \omega$).
It gives the fully transverse and Hall viscosities. Notice, that
the viscosities $\eta_0$, $\eta_1$ and $\eta_2$ are non-negative.
They determine the viscous dissipation (\ref{e:heat}) of motions of the
matter. In contrast, the Hall viscosities $\eta_3$ and $\eta_4$ may have
different sign (depending on the sign of electric charge of
momentum carriers) and do not contribute to the viscous dissipation
(\ref{e:heat}).

Fig.~\ref{fig:Hall} shows the dependence (\ref{eq:eta01234}) of all
the five viscosities on the Hall parameter $\xHall \propto B$.
At weak fields $B$, where $\xHall \ll 1$, the
electrons are non-magnetized (their collision frequency $1/\tau$ is
much higher than $\omega$). In this case the transverse viscosities
$\eta_1$ and $\eta_2$ converge to $\eta_0$,
and the Hall viscosities behave as $\eta_3\approx 2\xHall\eta_0$ and
$\eta_4\approx \xHall\eta_0$. The viscous stress tensor
$\sigma_{\alpha\beta}$ takes the form which is well-known in a
non-magnetized plasma \cite{LL_X}. In the case of moderately
magnetized electrons, $\xHall\sim 1$, all five viscosities are of
the same order of magnitude. At higher $B$, when $\xHall
\gg 1$, the electrons are strongly magnetized. Then $\eta_1\approx
\eta_0/(2\xHall)^2$, $\eta_2\approx \eta_0/\xHall^2$, $\eta_3
\approx \eta_0/(2\xHall)$ and $\eta_4\approx \eta_0/\xHall$. In this
regime the electrons frequently rotate about $\bm{B}$-lines and rarely
scatter off atomic nuclei.

\section{Viscosity in a neutron star crust}

The parameters of electrons and atomic nuclei in a neutron star
crust are provided by the models of the crust, e.g.
\cite{HPY2007}. For illustration, we take standard models
with one type of nuclei at any mass density $\rho$. We
will mainly use the model of the ground-state (cold catalyzed)
crust for the BSk21 equation of state approximated
analytically in \cite{BSk2013}.

We employ the effective electron relaxation time \cite{ChugYak2005}
which determines
the field-free shear viscosity $\eta_0$.
It seems to include the most elaborated physics
input. It is valid for gaseous, liquid and crystalline states of
atomic nuclei. It takes into account proper plasma screening of the
electron-nucleus interaction, multiphonon processes in the
crystalline phase in the harmonic lattice approximation, and finite
sizes of atomic nuclei. However, it neglects electron band structure
effects which can be important at sufficiently low temperatures.
If $B=0$, our consideration is valid at the same
conditions as in \cite{ChugYak2005} but at strong fields the
validity conditions are more restrictive (see below).

The electron relaxation time can be written as (e.g. \cite{ChugYak2005}),
\begin{equation}
\label{eq:tau} \tau = \frac{p_{\rm F}^2 v_{\rm F}}{12 \pi Z^2 e^4
n_{\rm i}\Lambda},
\end{equation}
where $n_{\rm i}$ is the number density of atomic nuclei and
$\Lambda$ is an effective Coulomb logarithm for electron-nucleus
scattering. It was analytically approximated in \cite{ChugYak2005}
using the method of an effective electron-nucleus potential first
implemented \cite{Pot1999} for the electron
conduction problem. The approximation is valid for a broad class
of spherical atomic nuclei which can be available in the neutron
star crust.

\begin{figure}
\includegraphics[width=\columnwidth]{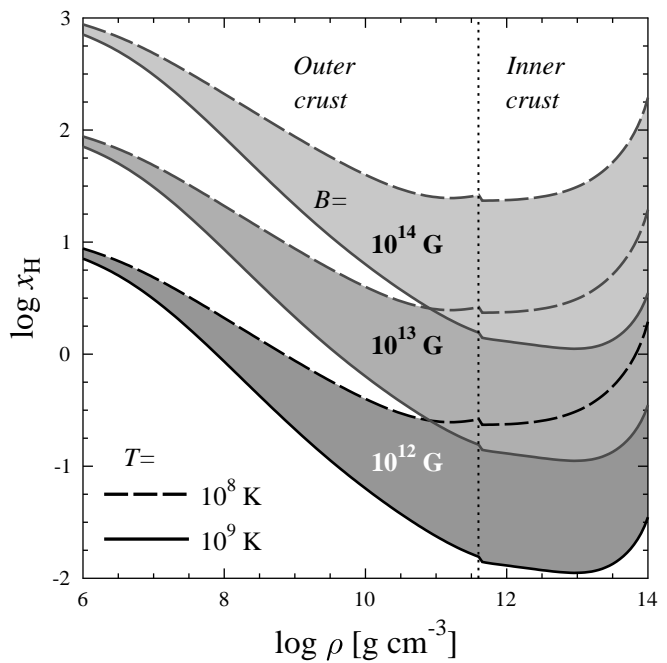}
\caption{ \label{fig:Hall} Hall parameter $\xHall$ versus mass
density $\rho$ in a neutron star crust for three values of the
magnetic field, $B=10^{12}$, $10^{13}$ and $10^{14}$~G, and two
temperatures, $T=10^8$ and $10^9$~K. When $B$ is fixed and the
temperature decreases from $10^9$ to $10^8$ K, an $\xHall(\rho)$
curve increases from an appropriate solid to a dashed line. The
vertical dotted line shows the neutron drip point. See text for
details. }
\end{figure}

Fig.~\ref{fig:Hall} shows the density dependence of the electron Hall
parameter $\xHall$ through the neutron star crust with the BSk21
equation of state. Small jumps of the nuclear composition due to
changes of nuclides with increasing density $\rho$ are smoothed out
as described in \cite{HPY2007,BSk2013}. The only jump which is left
occurs in the neutron-drip point ($\rho_\mathrm{ND}\approx 4.3 \times
10^{11}$ g~cm$^{-3}$) shown by the vertical dotted line. It
divides the neutron star crust into the outer and inner crust. The
inner crust disappears (transforms into the liquid core) at $\rho
\approx 1.4\times 10^{14}$ g~cm$^{-3}$, just after the highest
density displayed in Fig.\ \ref{fig:Hall}.

The displayed Hall parameter is a proper measure of magnetic effects
on the electron shear viscosity. It is shown for three values of
$B=10^{12}$, $10^{13}$, and $10^{14}$ G. Magnetic fields $B\sim
10^{12}-10^{13}$ G are typical for ordinary pulsars,
whereas higher $B$ are more typical for magnetars. We have taken
two temperatures, $T=10^9$ (solid lines) and $10^8$ K (dashed lines),
characteristic for ordinary young isolated (cooling) neutron stars
and magnetars. If $B$ is fixed and the temperature in the star falls
down from $10^9$ to $10^8$~K, the relaxation time
grows and an $\xHall(\rho)$ curve goes up from a
solid to a dashed line  amplifying
the electron magnetization. At the lowest displayed density
$\rho=10^6$ g~cm$^{-3}$ the relaxation time
is almost temperature independent,
so that the solid and dashed curves converge. The decrease of the
electron magnetization with increasing $\rho$ at fixed $B$ in the
outer crust is 
mainly provided 
by the decrease of the electron
gyrofrequency $\omega$ due to growing $\mu$. One sees that at
$B=10^{12}$ G the electrons stay weakly magnetized (in the given $T$
range) throughout the entire crust except for the surface layers. At
$B=10^{13}$ G, the electrons become strongly magnetized but only at
$\rho \lesssim 10^{10}$ g~cm$^{-3}$, whereas in the inner crust they are
moderately magnetized. If $B=10^{14}$ G, the electrons
in the outer crust are mostly strongly
magnetized, while in the inner crust they are strongly magnetized only at
$T\lesssim 10^8$ K. As expected, the strongest magnetization takes
place at the lowest densities.

\begin{figure*}
\includegraphics[width=\textwidth]{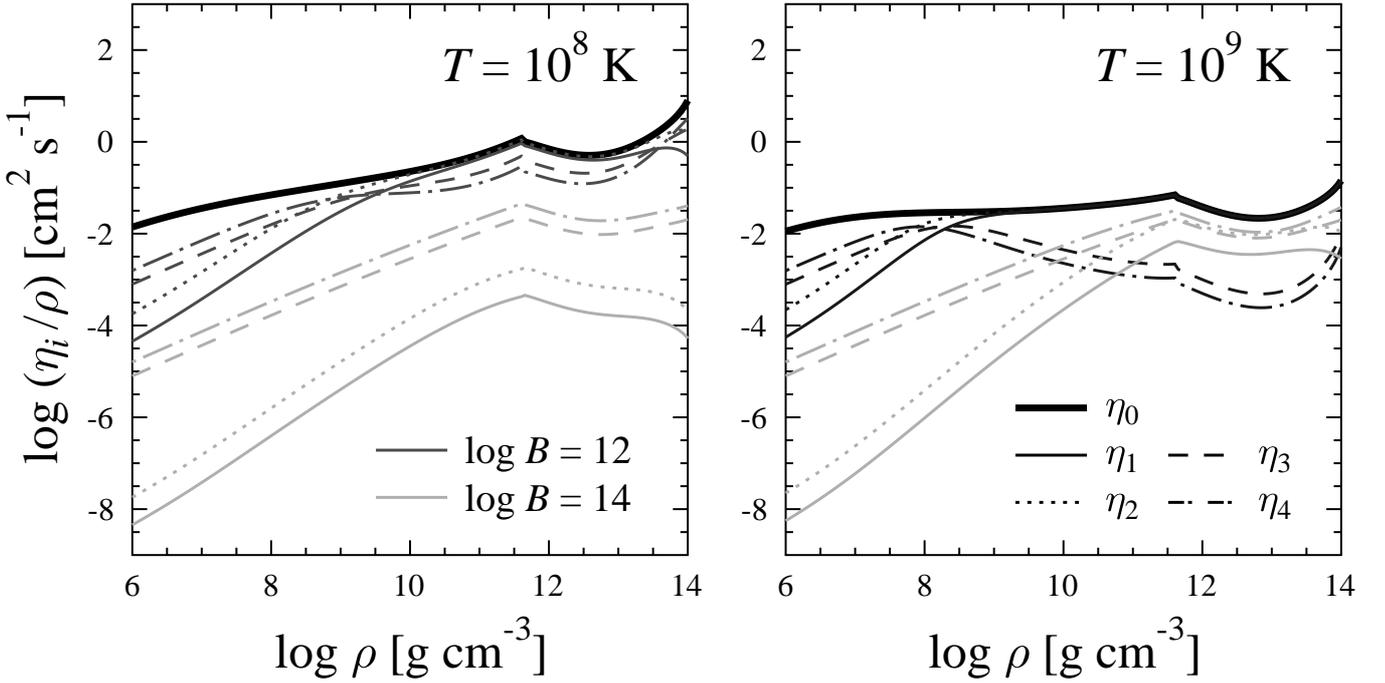}
\caption{\label{fig:viski} Density dependence of kinematic electron
shear viscosities $\eta_i/\rho$ ($i=0,\ldots,4$) for two values of
$B=10^{12}$ and $10^{14}$ G at two temperatures $T=10^8$~K (left)
and $10^9$~K (right). See text for details. }
\end{figure*}

Fig.\ \ref{fig:viski} shows the density dependence
through the entire crust of all five
electron viscosities $\eta_0,\ldots,\eta_4$ (lines of different type
and darkness). Calculations are done in the same formalism as in
Fig.\ \ref{fig:Hall}. Because the main trend is that $\eta_0 \propto
\rho$, we show the kinematic viscosities $\eta_i/\rho$. The left
panel of Fig.\ \ref{fig:viski} corresponds to $T=10^8$ K, while the
right panel is for $T=10^9$ K. The thick black line in each panel
gives the field-free viscosity $\eta_0/\rho$. Moderately dark
thinner lines present the viscosities $\eta_1,\ldots,\eta_4$ for
$B=10^{12}$ G, whereas light lines display these viscosities for
$B=10^{14}$ G. The behavior of the viscosities can be easily
understood from Figs. \ref{fig:lgx-lgeta} and \ref{fig:Hall}. At
$B=10^{14}$ G the electron magnetization is generally strong and 
the transverse viscosity components are the lowest one. If $B=10^{12}$ G,  
the magnetization is weaker and the Hall
viscosities could be the lowest ones.

So far we have considered only the crust made of the
BSk21 equation of state. We have checked also the BSk20
and BSk19 equations of state,
taking analytical approximations from \cite{BSk2013}, and the smooth
composition model \cite{HPY2007} of the crust.
The results appear to be almost
the same.

As mentioned above, our consideration of the
transverse and Hall viscosities $\eta_1,\ldots,\eta_4$ is
limited by the relaxation time approximation of the collision
integral in the Boltzmann equation. This approximation is valid
\cite{ZIMAN} as long as typical electron energy
transfers in electron-nucleus
collisions are smaller than the thermal energy width
$\sim k_{\rm B}T$ of the electron Fermi level. If so, the
Wiedemann-Franz relation between the electric and thermal
conductivities of electrons is fulfilled, which can be used as the
validity condition. Therefore, we have used our conductivity code
\cite{Pot1999,Gned2001} and checked the Widemann-Franz relation
numerically. If $T \gtrsim 10^9$~K, this relation is reasonably
well fulfilled throughout the entire crust meaning that our
consideration is justified. At $T \sim 3 \times 10^8$ K, it is well
fulfilled for $\rho \lesssim 10^{13}$ $\gcc$ but
violated at higher $\rho$ within a
factor of about 3. At $T \sim 10^8$ K, it works
reasonably well for $\rho \lesssim 10^{12}$ $\gcc$, but violated
at higher $\rho$ within a factor of 10. Therefore, our consideration
is strictly justified at not too high densities and not too low
temperatures; otherwise it can be treated
as semi-quantitative. It is possible to improve
the theory by using proper treatment of inelastic
electron collisions but
it is beyond the scope of this work.

We have also neglected some other effects which might affect the
results. For instance, at temperatures $T\lesssim T_B=\hbar
\omega_B/k_{\rm B}$ ($\omega_B$ being the cyclotron
frequency of the nuclei) the
frequencies of phonons responsible for electron transport in
Coulomb crystals of atomic nuclei will be affected by
magnetic fields and change the relaxation time $\tau$. As seen, for
instance, from Fig.\ 3 in \cite{Ofe2014}, $T_B$ is much lower than
$T$ for the conditions considered above and the effect is
not important here.

Note that at $T \lesssim 3 \times 10^7$ K, along with the
electron-nucleus scattering, an additional mechanism of
electron scattering by charged impurities may become important.
It has been widely discussed for electron electric and
thermal conductivities, e.g. \cite{Pot1999}, and it was also discussed
for the field free shear viscosity \cite{ChugYak2005}. It is easily
incorporated into the present formalism by using the combined
relaxation time $\tau$ given by $\tau^{-1}=\tau_{\rm ei}^{-1}
+\tau_{\rm e-imp}^{-1}$, where $\tau_{\rm ei}$ refers to
electron-nucleus collisions considered above, and
$\tau_{\rm e-imp}$ refers to electron scattering by
impurities \cite{ChugYak2005}. We do not discuss this case
here because of the lack of space and because the number density and
electric charges of the impurities are largely unknown.
At very low temperatures the relaxation is fully dominated
by electron-impurity collisions, $\tau=\tau_{\rm e-imp}$. These
collisions are thought to be elastic \cite{FI1976,ChugYak2005}
and described in the relaxation time
approximation. At
intermediate temperatures, both mechanisms, the electron-nucleus and
electron-impurity collisions, are equally important and the
problem is complicated by the description of inelastic
electron-nucleus collisions.

In addition, electron-electron collisions can also contribute to the
shear viscosity. In the field-free case this contribution was
studied in \cite{SHTERN2008} and found to be rather unimportant.
To the best of our knowledge, the shear viscosity
due to electron-electron
collisions in a magnetized plasma of neutron stars has not been
studied in the literature.

\section{White dwarfs}

The above results are equally valid for degenerate cores of white
dwarfs. The density range there is lower ($\rho \lesssim 10^{10}$
$\gcc$) and the nuclear composition is more restrictive (mainly
$^4$He, $^{12}$C, $^{16}$O). In such a plasma the relaxation time
approximation for describing the electron nucleus collisions is
valid to lower temperatures, down to a {  few million} K for a
typical density $\rho \sim 10^6$ $\gcc$. At this density, in the
field $B=10^9$ G the electrons are weakly magnetized, so that
$\eta_0 \approx \eta_1 \approx \eta_2$, whereas the Hall viscosities
$\eta_3$ and $\eta_4$ are lower than $\eta_0$ by about one order of
magnitude. If $B=10^{10}$ G, the electrons become moderately
magnetized, and at $B=10^{11}$ G their magnetization becomes strong,
with $\eta_1$ lower than $\eta_0$ by more than two orders of
magnitude. These effects can affect viscous damping of oscillations
of magnetic white dwarfs \cite{SEISMICWDS,BFIELDWDS}.

\section{Discussion and conclusions}

We have calculated the electron shear viscosity due to collisions of
electrons with atomic nuclei in a magnetized neutron star
crust. Calculations are done for strongly degenerate
relativistic or non-relativistic electrons. The
collision integral in the Boltzmann equation is taken in
the relaxation time approximation, but the effective relaxation time
$\tau$ is taken from more advanced field-free calculations
\cite{ChugYak2005}. The shear viscosity in
a magnetic field $B$ is described by the five viscosity coefficients,
$\eta_0 -\eta_4$. In our approximation, $\eta_0$ is
independent of $B$ and plays the
role of the viscosity along $\bm{B}$. Other viscosities, transverse
to $\bm{B}$ ($\eta_1$ and $\eta_2$) and Hall ones
($\eta_3$ and $\eta_4$) do depend on $B$. All these
viscosities are presented in the form convenient for computing
using any realistic composition of the neutron star crust. The
viscosity $\eta_0$ is valid for the same conditions as in
\cite{ChugYak2005}; other viscosities are restricted by the
applicability of the relaxation time approximation.

The dependence of the shear viscosities on $B$ is determined by the
Hall parameter $\xHall$. Generally, the transverse and Hall
viscosities are lower than $\eta_0$. At high electron magnetization,
$\xHall \gg 1$, they become much lower than $\eta_0$. The
magnetization increases with lowering $\rho$ and $T$ (Fig.\
\ref{fig:Hall}). It is much easier to magnetize the electrons in the
outer neutron star crust than in the inner one. At $T \sim 10^9$ K
the electrons in the inner crust remain weakly or moderately
magnetized even by magnetars' fields $B \sim 10^{14}$ G.

The shear viscosity is important for modeling many phenomena in
neutron stars, particularly, the relaxation of differential rotation
to a rigid-body rotation or the damping of waves and oscillations of
the stars, including the damping of instabilities accompanied by the
emission of gravitational waves. Since the transverse and Hall
viscosities are lower than $\eta_0$, one can expect that the viscous
damping of oscillating motions in a magnetized neutron star crust
will not significantly exceed the viscous damping of similar motions
in a non-magnetic crust \cite{ChugYak2005}. The anisotropic shear
viscosity in magnetized matter may produce noticeably different
damping times for motions of different geometry and
orientation with respect to the magnetic field. The present
results are equally valid for magnetic white dwarfs.

We have analyzed the basic features of the electron shear viscosity
in a neutron star crust. However, there remain a number of problems
to be solved. First, it would be important to extend the
calculations to sufficiently high densities and low temperatures
where the electron-nucleus scattering becomes inelastic and the
relaxation time approximation breaks down. It would be good to
include into these calculations the electron band structure effects
in crystalline matter (in analogy to the electron conduction problem
\cite{BANDS}) and consider mixtures of different nuclei. All this is
feasible but requires a lot of {  effort}, a good project for
future work. Second, it would be important to study the case of very
strong magnetic fields ($T \lesssim T_B$, see above) which affect
the vibration properties of atomic nuclei in dense matter and introduce
the dependence of the  relaxation time on $B$. One may expect, that
in this case the viscosity coefficients will contain three different
relaxation times (one for $\eta_0$, the second for $\eta_1$ and
$\eta_3$, and the third one for $\eta_2$ and $\eta_4$). If the
bottom of the crust contains the layer of exotic nuclear clusters
(the so called nuclear pasta) the electron shear viscosity there
will be more complicated (in analogy with a more complicated
conductivity \cite{PASTA}). Finally, very strong magnetic fields can
quantize electron states into Landau orbitals, and the viscosity
coefficients may show quantum oscillations due to the population of new
Landau levels with increasing density (similarly to quantum
oscillations of electron conductivities \cite{OSCILL}).

In addition, one should take into account other sources of shear
viscosity in neutron star envelopes. In particular, the main
contribution in the surface layers of neutron stars (roughly at
$\rho \lesssim 10^4$ $\gcc$) comes from ions (atomic nuclei). This
viscosity for a non-degenerate weakly coupled plasma of electrons
and ions in a magnetic field was calculated in \cite{BRAGIN} with
similar results on the existence of five shear viscosities of ions.
However, physical conditions in the outer layers of neutron stars
are more complicated because the ions there can be strongly coupled
and partially ionized. The effect of strong coupling on the shear
viscosity of ions at $B=0$ has been studied (e.g. \cite{VH75,WB78})
but the entire problem has not been solved. In the inner crust of a
neutron star some contribution to the shear viscosity may come from
free neutrons which interact with the atomic nuclei and can be in
superfluid state. Moreover, in addition to the shear viscosity,
there is the bulk viscosity of dense matter which can be
associated with weak interactions and have entirely different
physical properties (in analogy with the bulk viscosity in neutron
star cores \cite{BULK}). All these problems are almost not
considered and constitute an open field of physical kinetic of
neutron stars.

\acknowledgments
We are grateful to Andrey Chugunov for useful
comments. The work was supported by the Russian Science Foundation,
grant 14-12-00316.


\end{document}